# Bir Yükseköğretim Kurumu Web Sitesinin Ziyaretçi Verisi Analizi

## Analysis of the Visitor Data of a Higher Education Institution Website

**Ömer AYDIN**

**ÖZ**

İnternetin insan yaşamında her alana tesir ettiği günümüz dünyasında, internet birçok alanda olduğu gibi kurumsal web sitelerinde de değişime neden olmuştur. Kurumsal web siteleri daha dinamik, etkileşim daha çok olanak sağlayan ve yeni teknolojilere daha uyumlu olmalıdır. Web sitesinin kullanıcı, arama motorları ve diğer cihazlarla etkileşiminin uzman kişilerce incelenmesi ve bu etkileşime yönelik geliştirme ve değişikliklerin yapılması sitenin daha erişilebilir, esnek ve uyumlu olmasını sağlayacaktır. Bu doğrultuda yapılan bu çalışmada bir yükseköğretim kurumuna ait web sitesinin verileri incelenmiştir. İnceleme için 2013 yılı ile 2019 yılı arasındaki toplanmış ziyaretçi verileri kullanılmıştır. Kapsamlı incelemelerin ve verilerin yer aldığı çalışmada, ziyaretçi trafik analizinden geliştirme önerilerine kadar önemli birçok noktaya yer verilmiştir. Özellikle sitenin mobil cihazlar ile uyumluluğu, resim ve videoların optimizasyonu, kullanıcıların demografik özellikleri, dil seçenekleri ile erişilen içeriğin zamana göre yoğunluğu analizleri ile faydalı bilgilere ulaşılmıştır.

**Anahtar Sözcükler:** Trafik analizi, Ziyaretçi analizi, Web sitesi analizi, Kurumsal web sitesi

**ABSTRACT**

In today's world, the internet affects every aspect of human life; it has caused changes in corporate websites as well as in many other areas. Corporate websites should be more dynamic, more interactive, and more compatible with new technologies. The interaction of the web site with users, search engines, and other devices has to be examined by experts, and improvements and changes should be made for this interaction. In this study, a higher education institution web site was examined. Visitor data collected between 2013 and 2019 were used for the analysis. In the study, which includes a wide range of examinations and data, important findings from traffic analysis to development suggestions were included. In particular, useful information has been obtained through the compatibility of the site with mobile devices, optimization of pictures and videos, geographical features of users, language options, and density analysis of the content accessed over time.

## GİRİŞ

İnternetin temelleri 1969 yılında Amerika Birleşik Devletlerinde askeri amaçlarla kurulan ve adına ARPANET denilen bir ağ ile atılmıştır. ARPANET'in yerini 1980'li yılların başında Ulusal Bilim Vakfı (National Science Foundation) tarafından ortaya koyulan NSFNET almıştır. 1980 yılında genel kullanıma açılan bu ağ ile birlikte kullanım yaygınlaşmıştır. 1990 yılına gelindiğinde çoğunluğunun Amerika'da yer aldığı ağların hepsi birleştirilerek internet meydana gelmiştir (Küçük, 2017). Önceleri Dial-up vb. düşük hızla bağlantı sağlayan ve mevcut telefon hatlarını kullanan internet bağlantısı imkânları bulunuyordu. 2000'li yıllara gelindiğinde ülkemizde ADSL teknolojisi kullanılmaya başlandı (Businessht Bloomberght, 2015). Kısa sürede ADSL yaygınlaşmaya başladı. ADSL yaygınlaşırken aynı zamanda veri aktarım

Ömer AYDIN (✉)
ORCID ID: 0000-0002-7137-4881

Dokuz Eylül Üniversitesi, İktisadi ve İdari Bilimler Fakültesi, İzmir, Türkiye
*Dokuz Eylül University, Faculty of Economics and Administrative Sciences, İzmir, Turkey*
omer.aydin@deu.edu.tr

hızı bakımından da gelişmeye devam etti. Son kullanıcılara fiber altyapılar veya uydu sistemleri kullanılarak yüksek hızda internet hizmeti de verilmeye başlandı. Bu teknolojilere paralel olarak kablosuz ve mobil ağlar da gelişmesini sürdürüyordu. Özellikle cep telefonu alanındaki donanımsal gelişmeler ve erişilebilirliğin artması ile mobil ağ teknolojilerindeki gelişmeler daha fazla insana ulaşmaya başladı. Günümüzde video, müzik, resim ve diğer içerikler kablolu veya kablosuz bağlantlar vasıtası ile yüksek hızda kullanıcılara ulaştırılabilmektedir. Öyle ki, 4G bağlantısı ile saniyede 100 Megabit (Mbps) hıza, 5G bağlantısı ile saniyede 10 Gigabit (Gbps) hıza ulaşılabilmektedir (Kavanagh, 2019). 5G bağlantısı ile sağlanan veri aktarım hızı 4G ile sağlanabilecek hızın yaklaşık 100 katıdır. Aynı şekilde fiber bağlantı ile saniyede 100 Terabit (Tbps) hıza ulaşılmıştır (Shiftdelete.net, 2011). Uydu sistemleri ile saniyede 506 Megabit (Mbps) hıza ulaşmak mümkün hâle gelmiştir (Wikipedia, 2018).

Tablo 1'de 2019 yılı Aralık verilerine göre internet kullanımı bakımından en çok kullanıcıya sahip ilk 20 ülke verilmiştir. Şekilde görülebileceği üzere 1 milyardan fazla insan nüfusuna sahip Çin 854 milyon internet kullanıcısı ile dünyada lider konumdadır. Nüfus ile internete erişim oranı bakımından değerlendirildiğinde ise %100 ile Vietnam, %94.66 oran ile ABD ve %94.44 oranla Almanya başı çekmektedir. 2000 yılı ile 2020 yılı arasındaki gelişim bakımından değerlendirildiğinde ise %94099 ile Bangladeş internetin en çok geliştiği ülke olmuştur. Bangladeş'i Nijerya, Vietnam ve İran izlemektedir. Türkiye ise 2020 yılı itibarı ile 69 milyon internet kullanıcına ulaşmış ve nüfusunun %81.94'ü internete erişen bir ülkedir. Türkiye'de internete erişim 2000 yılı ile 2020 yılı arasında karşılaştırıldığında da %3355 oranında artmıştır.

Web siteleri internet kullanımının en yoğun olarak kullanıldığı ortamlardandır. İnternet altyapısını kullanan uygulamalar, oyunlar vb. birçok ortam bulunmasına rağmen kullanıcılar bilgiye internet sayfaları vasıtası ile ulaşmaya devam etmektedir (Sohn, Li, Griswold, & Hollan, 2008). Bu nedenle internet sayfalarının mevcut teknolojik gelişmelere ayak uydurması, kullanıcılara kullanım kolaylığı sunması ciddi önem kazanmaktadır. Özellikle kurumsal web sitelerinde sunulan bilginin kullanıcıya ulaşması için ilgili web sitelerinin analizlerinin yapılması, kullanıcı taleplerine ve davranışlarına uygun değişikliklerin yapılması büyük önem taşımaktadır. Bununla birlikte arama motoru

**Tablo 1:** En Yüksek İnternet Kullanıcısına Sahip 20 Ülke (Internet World Stats, 2020)

| # | Ülke veya Bölge | İnternet Kullanıcısı Sayısı | | Nüfus | | İnternet Artışı 2000-2020 (%) | İnternet Kullanıcısı/Nüfus (2020) (%) |
|---|---|---|---|---|---|---|---|
| | | 2020 Yılı | 2000 Yılı | 2020 (Hesaplanan) | 2000 Yılı | | |
| 1 | Çin | 854000000 | 22500000 | 1439062022 | 1283198970 | 3695 | 59.34 |
| 2 | Hindistan | 560000000 | 5000000 | 1368737513 | 1053050912 | 11100 | 40.91 |
| 3 | ABD | 313322868 | 95354000 | 331002651 | 281982778 | 229 | 94.66 |
| 4 | Endonezya | 171260000 | 2000000 | 273523615 | 211540429 | 8463 | 62.61 |
| 5 | Brezilya | 149057635 | 5000000 | 212392717 | 175287587 | 2881 | 70.18 |
| 6 | Nijerya | 126078999 | 200000 | 206139589 | 123486615 | 62939 | 61.16 |
| 7 | Japonya | 118626672 | 47080000 | 126854745 | 127533934 | 152 | 93.51 |
| 8 | Rusya | 116353942 | 3100000 | 145934462 | 146396514 | 3653 | 79.73 |
| 9 | Bangladeş | 94199000 | 100000 | 164689383 | 131581243 | 94099 | 57.20 |
| 10 | Meksika | 88000000 | 2712400 | 132328015 | 2712400 | 3144 | 66.50 |
| 11 | Almanya | 79127551 | 24000000 | 83783942 | 81487757 | 230 | 94.44 |
| 12 | Filipinler | 79000000 | 2000000 | 109581078 | 77991569 | 3850 | 72.09 |
| 13 | Türkiye | 69107183 | 2000000 | 84339067 | 63240121 | 3355 | 81.94 |
| 14 | Vietnam | 68541344 | 200000 | 68541344 | 200000 | 34171 | 100.00 |
| 15 | Birleşik Krallık | 63544106 | 15400000 | 67886011 | 58950848 | 313 | 93.60 |
| 16 | İran | 67602731 | 250000 | 83992949 | 66131854 | 26941 | 80.49 |
| 17 | Fransa | 60421689 | 8500000 | 65273511 | 59608201 | 611 | 92.57 |
| 18 | Tayland | 57000000 | 2300000 | 69799978 | 62958021 | 2378 | 81.66 |
| 19 | İtalya | 54798299 | 13200000 | 60461826 | 57293721 | 315 | 90.63 |
| 20 | Mısır | 49231493 | 450000 | 102334404 | 69905988 | 10840 | 48.11 |
| | En Üstteki 20 Ülke | 3241273512 | 251346400 | 5233377837 | 4312497691 | 1189 | 61.93 |
| | Dünyanın Geri Kalan Ülkeleri | 1332876622 | 109639092 | 2563237873 | 1832509298 | 1116 | 52.00 |
| | **Toplam (Dünya)** | **4574150134** | **360985492** | **7796615710** | **6145006989** | **1167** | **58.67** |

optimizasyonu olarak bilinen ve web sitesinin veya içeriğinin arandığında arama motoru sonuçları arasında daha ön sıralarda yer almasını sağlamak için yapılan işlemler için de bu bilgiler önem taşımaktadır.

Bu çalışmada kurumsal bir web sitesi olarak ülkemiz yükseköğretim kurumlarından birine ait web sitesinin 2013 yılından itibaren kayıt altına alınan ziyaretçi trafik bilgileri incelenmiştir. Çalışmanın yapılma amacı elde edilen bulgular ile web sitenin performansının ve erişilebilirliğinin artırılması ve yeni teknolojilere uyumunun sağlanması için yol göstermesidir. Sitenin kaynak (sunucu, bağlantı hızı ve yazılım gibi) ihtiyaçlarının planlanması ve yönetilmesi için bulgular elde etmek bu çalışmanın hedefleri arasında yer alır. Buradan elde edilen veriler ve yapılan analizler ziyaretçilerin dil, ülke, cihaz, bağlantı operatör vb. özelliklerini de görebilme imkânı sağladığından dolayı kullanıcı tercihlerini ve kullandıkları teknolojileri değerlendirerek kullanım kolaylığını artırmaya yönelik geliştirmeler yapılma imkânı sunmaktadır.

Bu çalışma ile aşağıda listelenen bazı sorulara cevap aranmaktadır.

- Web siteleri kullanıcılar tarafından nasıl bulunur?
- Ziyaretçiler tarafından kullanılan içerik nedir?
- Kullanıcılar web sitesine ne sıklıkta geri geliyorlar (ve kaç tane yeni kullanıcı var)?
- Kullanıcılar hakkında ne biliyoruz?
- Ziyaretçilerin yaşadığı ülkeler ile siteyi görüntülediklerindeki varsayılan dil tercihi hangisidir?
- Ziyaretçilerin hangi saat dilimlerinde ve yılın hangi dönemlerinde ziyaretleri yoğunlaşıyor?
- Web sitemizi ziyaret etmek için hangi cihazlar kullanıldı?
- Web sitemize erişimde hangi ekran çözünürlükleri, dil, tarayıcı ve servis sağlayıcı kullanıldı?

Çalışma 2000'li yılların başından beri çevrimiçi olarak hizmet vermekte olan kurumsal bir web sitesi kullanılarak yapılmıştır. İlgili web sitesi üzerinden 2013-2019 yılları arasında 6 yıllık toplanan veriler çalışmanın verisini oluşturmaktadır. Belirtilen zaman dilimine ait iki milyona yakın kullanıcı girişi ve 8.5 milyonun üzerinde oturum ile oluşan demografik, sistem ve zaman bilgilerinden oluşan bir veri seti bulunmaktadır.

## VERİ ve YÖNTEM

Kurumsal bir internet sitesinin incelenmesinde dikkat edilmesi gereken bazı temel unsurları aşağıdaki gibi sıralayabiliriz:

- Sitenin tasarımının ve grafik arayüzünün incelenmesi (İlgili arayüzün kurumsal kimlikle uygunluğu),
- İçerik olarak doğruluk ve yeterliliğin incelenmesi,
- Web sitesi için kullanılan kaynak yönetim sitemi vb. yazılımların altyapısının incelenmesi,
- Web sitesi hızının, kullanılabilirlik ve işlevselliğinin incelenmesi,

- Arama motorları bakımından inceleme yapılması,
- Web sitesi adresi ve sunucu hizmeti bakımından incelenmesi,
- Web sitesinin gösteriminin ve erişiminin sağlandığı bilgisayar, tablet, mobil cihazlar vb. uyumluluk analizi,
- Web sitesinin gösteriminin sağlandığı internet tarayıcıları bakımından uyumluluğunun incelenmesi,
- Sosyal medya ortamları ile uyumunun ve iletişiminin incelenmesi (Arma Digital, 2019).

Bu ve bunun gibi sayısı artırılabilecek birçok maddede belirtilen ölçütler bakımından web sitelerinin analizinin ve incelenmesinin yapılması mümkündür. Günümüzde web sitesi trafik analizi için birçok araç kullanılabilmektedir. Özellikle web sitesi yazılımları içine yerleştirilen Javascript tabanlı kodlar vasıtası ile kullanıcı davranışlarının ve bilgilerinin toplanması mümkün olmaktadır. Bu şekilde bilgi toplayan birçok uygulamaya internet üzerinden ücretli veya ücretsiz olarak ulaşmak mümkündür.

## BULGULAR

Ziyaretçi verilerinden elde edilen bulguları bu başlık altında görmek mümkündür. Her bir bulgu hakkındaki bilgiler tek tek detayları ile verilmiştir.

Şekil 1'de, 10 Ekim 2013 tarihi ile 2 Eylül 2019 tarihleri arasındaki siteye erişen tekil kullanıcı sayılarının grafiği görülmektedir. Bu şekilde genel hatları ile ziyaretçi sayılarının nasıl dağıldığı görülebilmektedir. İlk gün yani 10 Ekim 2013 tarihinde 592 kullanıcı siteyi ziyaret etmiştir. 1 Temmuz 2015 tarihinde ise 9061 tekil kullanıcı ile tüm zamanların en yüksek erişim sayısına ulaşılmıştır.

**Tablo 2:** 5000 ve Üzeri Kullanıcının Erişimi Olan Günlerin Aylara Göre Dağılımı

| Ay | 5000 üzerinde erişim olan gün sayısı |
|---|---|
| Haziran | 29 |
| Ekim | 21 |
| Ocak | 21 |
| Kasım | 18 |
| Temmuz | 17 |
| Eylül | 15 |
| Şubat | 14 |
| Mart | 12 |
| Nisan | 11 |
| Ağustos | 9 |
| Aralık | 6 |
| Mayıs | 5 |
| **Toplam** | **178** |

Şekil 1'de 5000 ve üzeri tekil kullanıcı ziyaretinin olduğu günlerin tepe noktaları olduğu görülebilmektedir. Bu nedenle bu parametreyi belirleyici bir sınır olarak seçerek gruplama yapıl-

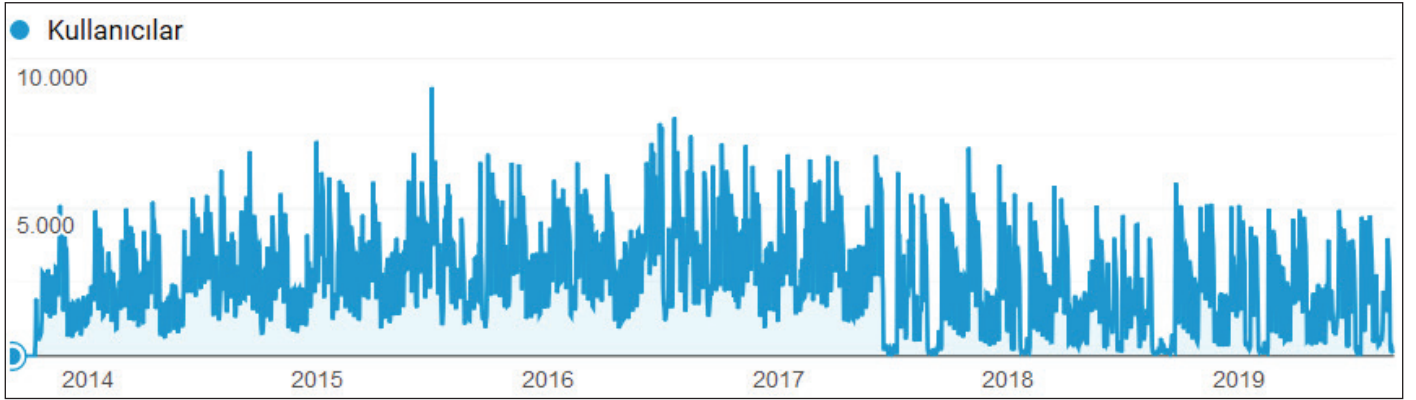

**Şekil 1:** 2013 ve 2019 yılları arasında tekil kullanıcı trafik dağılımı grafiği.

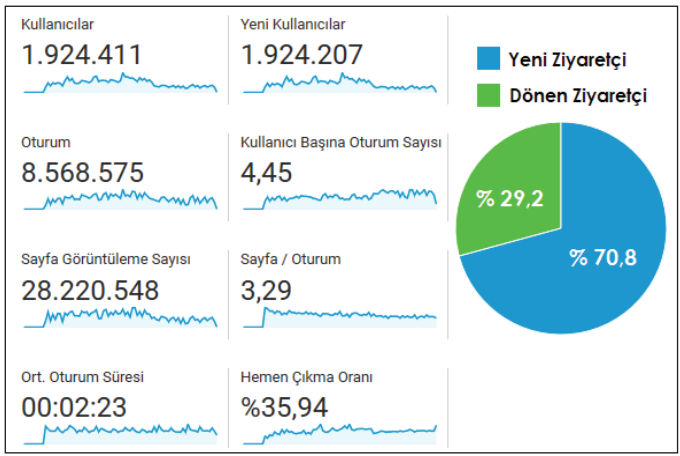

**Şekil 2:** Site ziyaretçi genel bilgileri.

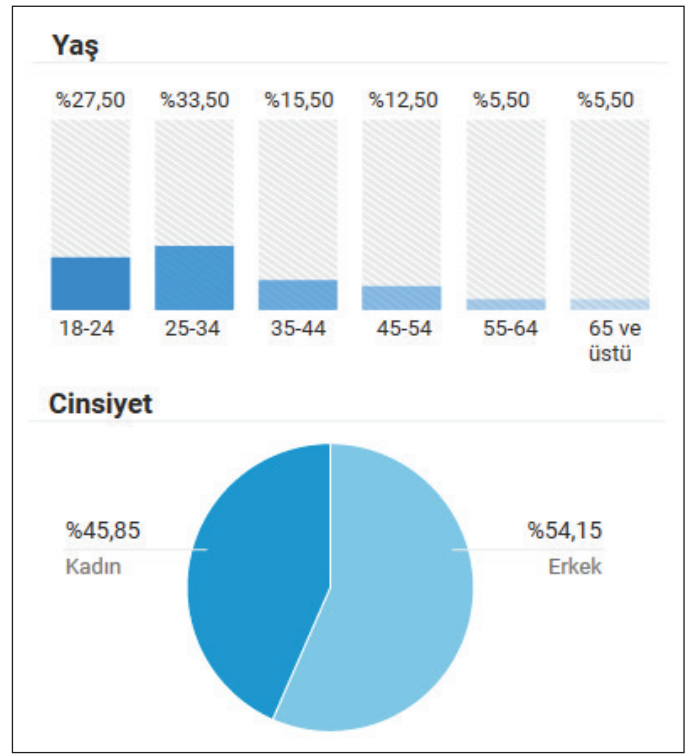

**Şekil 3:** Ziyaretçilerin yaş ve cinsiyet bakımından dağılımı.

dığında ise Tablo 2'de verilmiş bilgilere ulaşılabilmektedir. Bu tablodaki veriler görüldüğü üzere 5000 kullanıcı ve üzeri erişim olan gün sayısına bakıldığında Mayıs, Aralık ve Ağustos aylarında en düşük sayıda veri bulunurken Ekim, Ocak ve Haziran ayları en çok 5000 üzeri kullanıcı erişimi olan gün sayısına sahip aylar olmuştur.

Şekil 2'de 10 Ekim 2013 ve 2 Eylül 2019 tarihleri arasında siteyi ziyaret eden tekil kullanıcı sayı, oturum sayısı, kullanıcı başına oturum sayısı, sayfa görüntüleme sayısı, sayfa başına oturum sayısı, ortalama oturum süresi, hemen çıkma oranı ve yeni kullanıcı ve tekrar gelen kullanıcılara ait bilgilere yer verilmiştir. İlgili şekilden görülebileceği gibi iki milyona yakın tekil kullanıcı ziyareti ve 8.5 milyonun üstünde oturum gerçekleşmiştir. Her bir kullanıcı yaklaşık 4.5 oturum ile siteye erişmiştir. Buradaki oturum bir kullanıcının farklı zamanlarda siteyi ziyaret etmesi anlamına gelmektedir. Bir kullanıcının sitede ortalama olarak 2 dakika 23 saniye kaldığı hesaplanmıştır. Siteye giren 100 kullanıcında yaklaşık 36'sının ise hemen çıktığı görülebilmektedir. Sitenin ziyaretçilerinin yaklaşık %70'i yeni kullanıcılar, geriye kalan %29'u ise daha önce siteyi ziyaret edip tekrar siteye giren kullanıcılardan oluşmaktadır.

Şekil 3'te ise ilgili tarih aralığında siteyi ziyaret etmiş kullanıcıların yaş ve cinsiyet bakımından oranları verilmektedir. Belirtilen dönemde siteyi ziyaret eden toplam kullanıcının yaklaşık %55'i erkek, %45'i kadın kullanıcılardır. Aynı tarih aralığındaki toplam ziyaretçilerin %27.5'i 18-24 yaş, %33.5'i 25-34 yaş, %15.5'i 35-44 yaş ve geri kalan %23.5'lik kısmı ise 45 yaş ve üstü kullanıcılardan oluşmaktadır. Kısacası kullanıcıların %61'i 35 yaş altındadır.

Şekil 4 ve Şekil 5'teki grafik ve tablolarda dil ve ülkeye göre ziyaretçi dağılımı verilmiştir. Şekil 4'teki dil bilgisi tarayıcı verileri incelenerek elde edilen bilgilerden gelmektedir. Detaylı göz atıldığında oturumların yaklaşık %95'inin Türkçe ile yaklaşık %4.5'inin İngilizce ve geri kalan kısmını Rusça, Almanca gibi diller oluşturmaktadır.

| | Dil | Yeni Kullanıcılar | | Oturum | |
|---|---|---|---|---|---|
| | | **1.927.495** | | **8.568.575** | |
| | | Toplam Yüzdesi: %100,17 (1.924.207) | | Toplam Yüzdesi: %100,00 (8.568.575) | |
| 1. | tr | 1.009.253 | (%52,36) | 4.150.725 | (%48,44) |
| 2. | tr-tr | 788.796 | (%40,92) | 3.966.928 | (%46,30) |
| 3. | en-us | 57.039 | (%2,96) | 192.332 | (%2,24) |
| 4. | en | 41.371 | (%2,15) | 173.075 | (%2,02) |
| 5. | en-gb | 7.834 | (%0,41) | 41.959 | (%0,49) |
| 6. | ru | 5.727 | (%0,30) | 10.486 | (%0,12) |
| 7. | (not set) | 4.467 | (%0,23) | 4.732 | (%0,06) |
| 8. | de-de | 1.473 | (%0,08) | 3.967 | (%0,05) |
| 9. | ru-ru | 1.902 | (%0,10) | 3.698 | (%0,04) |
| 10. | c | 3.304 | (%0,17) | 3.334 | (%0,04) |

**Şekil 4:** Dile göre ziyaretçi dağılımı.

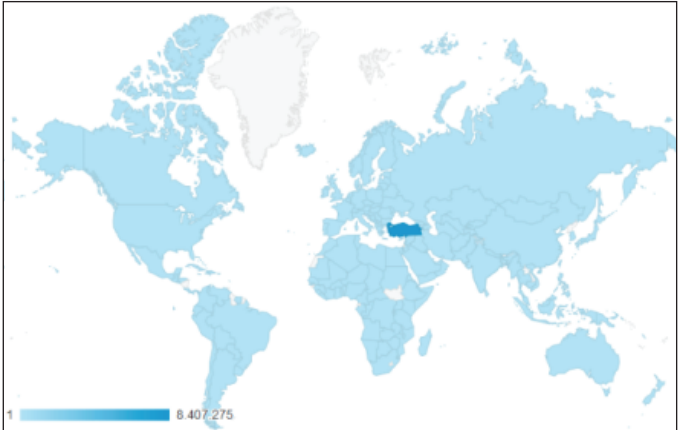

| | Ülke | Yeni Kullanıcılar | | Oturum | |
|---|---|---|---|---|---|
| | | **1.927.495** | | **8.568.575** | |
| | | Toplam Yüzdesi: %100,17 (1.924.207) | | Toplam Yüzdesi: %100,00 (8.568.575) | |
| 1. | Turkey | **1.868.803** | (%96,96) | 8.407.275 | (%98,12) |
| 2. | United States | 12.752 | (%0,66) | 38.170 | (%0,45) |
| 3. | Russia | 7.948 | (%0,41) | 9.995 | (%0,12) |
| 4. | (not set) | 6.556 | (%0,34) | 20.676 | (%0,24) |
| 5. | Germany | 4.212 | (%0,22) | 11.374 | (%0,13) |
| 6. | United Kingdom | 3.299 | (%0,17) | 9.999 | (%0,12) |
| 7. | Cyprus | 2.680 | (%0,14) | 5.001 | (%0,06) |
| 8. | India | 2.366 | (%0,12) | 12.514 | (%0,15) |
| 9. | Indonesia | 1.937 | (%0,10) | 12.591 | (%0,15) |
| 10. | Azerbaijan | 1.853 | (%0,10) | 2.504 | (%0,03) |

**Şekil 5:** Dil ve ülkeye göre ziyaretçi dağılımı.

Şekil 5'in üst kısmında yer alan dünya ülkeler haritasına bakıldığında çok geniş bir coğrafyadan sayfaya ziyaretçi geldiği görülmektedir. Neredeyse dünya üzerindeki tüm ülkelerden siteye giriş gerçekleşmiştir. Şeklin alt kısmında yer alan bilgilere bakıldığında ise siteye giriş yapan kullanıcıların ülkelerinden ilk 10'u görülebilmektedir. Türkiye %98 ile birinci sırada yer alır-

| | Kıta | Yeni Kullanıcılar | | Oturum | |
|---|---|---|---|---|---|
| | | **1.927.495** | | **8.568.575** | |
| 1. | Asia | 1.882.126 | (%97,65) | 8.449.480 | (%98,61) |
| 2. | Europe | 23.562 | (%1,22) | 54.981 | (%0,64) |
| 3. | Americas | 13.803 | (%0,72) | 40.370 | (%0,47) |
| 4. | (not set) | 6.556 | (%0,34) | 20.676 | (%0,24) |
| 5. | Africa | 1.298 | (%0,07) | 2.844 | (%0,03) |
| 6. | Oceania | 150 | (%0,01) | 224 | (%0,00) |

| | Alt Kıta | Yeni Kullanıcılar | | Oturum | |
|---|---|---|---|---|---|
| | | **1.927.495** | | **8.568.575** | |
| 1. | Western Asia | 1.874.894 | (%97,27) | 8.419.694 | (%98,26) |
| 2. | Northern America | 13.072 | (%0,68) | 38.990 | (%0,46) |
| 3. | Western Europe | 7.960 | (%0,41) | 23.920 | (%0,28) |
| 4. | (not set) | 6.556 | (%0,34) | 20.676 | (%0,24) |
| 5. | Eastern Europe | 9.863 | (%0,51) | 14.932 | (%0,17) |
| 6. | Southern Asia | 3.116 | (%0,16) | 13.950 | (%0,16) |
| 7. | Southeast Asia | 2.381 | (%0,12) | 13.451 | (%0,16) |
| 8. | Northern Europe | 3.886 | (%0,20) | 11.204 | (%0,13) |
| 9. | Southern Europe | 1.853 | (%0,10) | 4.925 | (%0,06) |
| 10. | Eastern Asia | 1.179 | (%0,06) | 1.550 | (%0,02) |

**Şekil 6:** Kıta ve bölgeye göre ziyaretçi dağılımı.

| | Şehir | Yeni Kullanıcılar | | Oturum | |
|---|---|---|---|---|---|
| | | **1.927.495** | | **8.568.575** | |
| 1. | Izmir | 1.133.379 | (%58,80) | 6.360.939 | (%74,24) |
| 2. | Istanbul | 223.894 | (%11,62) | 802.264 | (%9,36) |
| 3. | Ankara | 127.385 | (%6,61) | 386.611 | (%4,51) |
| 4. | Antalya | 28.841 | (%1,50) | 92.827 | (%1,08) |
| 5. | Adana | 29.731 | (%1,54) | 68.462 | (%0,80) |
| 6. | Bursa | 21.069 | (%1,09) | 51.338 | (%0,60) |
| 7. | (not set) | 19.124 | (%0,99) | 50.716 | (%0,59) |
| 8. | (not set) | 12.686 | (%0,66) | 48.591 | (%0,57) |
| 9. | Manisa | 15.755 | (%0,82) | 44.018 | (%0,51) |
| 10. | Aydın | 13.803 | (%0,72) | 36.330 | (%0,42) |

**Şekil 7:** Ziyaretçilerin şehir bazlı dağılımı.

ken onu Amerika Birleşik Devletleri, Rusya, Almanya ve Birleşik Krallık izlemektedir.

Şekil 4 ve Şekil 5'teki dil ve ülke bilgilerine ek olarak Şekil 6'da kıtalara ve bu kıtalara bölgelerine göre ziyaretçi ve oturum sayıları görülebilmektedir. Toplam oturumun yaklaşık %98'i Asya, %64'ü Avrupa ve diğer kısımları Amerika, Afrika ve Avustralya kıtası olmak üzere sıralanmıştır. Bölge olarak inceleyecek olursak toplam oturumun %98'i Batı Asya, %46'sı Kuzey Amerika ve geri kalanı sırası ile Batı Avrupa, Doğu Avrupa, Güney Asya, Kuzey Avrupa şeklinde devam etmektedir.

Sitenin ilgili tarih aralığında hangi şehirlerden ziyaret edildiği bilgisine Şekil 7'de verilmiştir. Şekil 7'ye bakıldığında en çok ziyaretçi %58 ve oturum %74 ile İzmir'den gerçekleşmiştir. Oturum sayısı bakımından bu şehri %9.36 ile İstanbul, %4.51 ile Ankara ve %1.08 ile Antalya izlemektedir.

Ayrıca tüm şehirlerin listesinden sadece yurtdışındaki şehirleri içeren ikinci bir liste oluşturularak Şekil 8'de verilmiştir. Şekil 8'e baktığımızda ise oturum sayısı bakımından genel sıralamada 26. sırada yer alan New Delhi %0.12 lik oturum oranı ile yurtdışı şehirleri arasında birinci sırada yer almaktadır. Onu %0.08 ile Moskova ve %0.05 ile Londra izlemektedir.

| | Şehir | Yeni Kullanıcılar | Oturum |
|---|---|---|---|
| | | Toplam (1.924.207) | Toplam (8.568.575) |
| 26. | New Delhi | 1.548(%0,08) | 10.558(%0,12) |
| 34. | Moscow | 5.611(%0,29) | 6.939(%0,08) |
| 46. | London | 1.455(%0,08) | 4.402(%0,05) |
| 53. | Frankfurt | 585(%0,03) | 3.543(%0,04) |
| 61. | New York | 1.125(%0,06) | 3.310(%0,04) |
| 74. | Dubai | 501(%0,03) | 2.062(%0,02) |
| 76. | Amsterdam | 521(%0,03) | 1.969(%0,02) |
| 81. | Baku | 1.292(%0,07) | 1.724(%0,02) |
| 85. | Berlin | 638(%0,03) | 1.527(%0,02) |
| 86. | Kyiv | 313(%0,02) | 1.444(%0,02) |
| 87. | Nicosia | 805(%0,04) | 1.403(%0,02) |
| 92. | Coffeyville | 361(%0,02) | 1.365(%0,02) |
| 93. | Slough | 203(%0,01) | 1.319(%0,02) |
| 95. | Paris | 185(%0,01) | 1.059(%0,01) |
| 101. | Pune | 160(%0,01) | 982(%0,01) |
| 102. | Lynn | 274(%0,01) | 948(%0,01) |
| 103. | Jakarta | 268(%0,01) | 929(%0,01) |
| 106. | Mountain View | 744(%0,04) | 848(%0,01) |
| 111. | Pavia | 24(%0,00) | 700(%0,01) |
| 112. | Granada | 115(%0,01) | 699(%0,01) |
| 113. | Saint Petersburg | 400(%0,02) | 679(%0,01) |
| 114. | Amsterdam-Zuidoost | 131(%0,01) | 642(%0,01) |
| 116. | Chicago | 239(%0,01) | 588(%0,01) |
| 117. | Lagos | 101(%0,01) | 556(%0,01) |

**Şekil 8:** Yurtdışı ziyaretçilerinin şehirlere göre dağılımı.

| | Tarayıcı | Yeni Kullanıcılar | Oturum |
|---|---|---|---|
| | | 1.927.495 | 8.568.575 |
| 1. | Chrome | 1.127.771 (%58,51) | 5.097.554 (%59,49) |
| 2. | Safari | 499.567 (%25,92) | 2.254.936 (%26,32) |
| 3. | Android Browser | 67.196 (%3,49) | 502.489 (%5,86) |
| 4. | Internet Explorer | 80.919 (%4,20) | 185.443 (%2,16) |
| 5. | Firefox | 51.219 (%2,66) | 125.700 (%1,47) |
| 6. | Samsung Internet | 7.246 (%0,38) | 101.357 (%1,18) |
| 7. | YaBrowser | 19.294 (%1,00) | 72.884 (%0,85) |
| 8. | Safari (in-app) | 14.665 (%0,76) | 51.246 (%0,60) |
| 9. | Opera Mini | 8.199 (%0,43) | 46.439 (%0,54) |
| 10. | Opera | 12.279 (%0,64) | 41.482 (%0,48) |

**Şekil 9:** İnternet tarayıcısına göre ziyaretçi sayıları.

Şekil 9'da, kullanıcıların siteyi ziyaretlerinde esnasında hangi internet tarayıcısını kullandıkları bilgisi yer almaktadır. Detaylı incelendiğinde kullanıcıların yaklaşık %58'i ve oturumların yaklaşık %59'unun Chrome tarayıcısını kullandığı görülmektedir. Bu tarayıcıyı yaklaşık %25 ile Safari izlemektedir. Diğer tarayıcılar ise Andorid browser, Internet Explorer ve Firefox, Samsung® Internet, YaBrowser, Safari(in-app), Opera Mini ve Opera. şeklindedir.

Şekil 10 üzerinde kullanıcıların ziyaret esnasında hangi işletim sistemi üzerinden erişim sağladıkları bilgisi yer almaktadır. Bu işletim sistemleri bakımından incelendiğinde ise oturumların %46, kullanıcıların ise %28'inin Android kullanıcısı olduğu görülmektedir. Bu oranları %28 kullanıcı ve %29 oturum oranı ile IOS izlemektedir. Üçüncü sırada ise %40 kullanıcı, %22 oturum oranı ile Windows işletim sistemi yer almaktadır.

Şekil 11'de ise erişim sağlanan cihazların kategori bakımından gruplaması yapılmıştır. Bu gruplamaya göre cep telefonu vb. (mobil cihazlar) üzerinden erişim oranı kullanıcı bakımından %57, oturum bakımından %76 olarak gerçekleşmiştir. Masaüstü olarak tabir edilen bilgisayarlardan erişim oranı kullanıcı bakımından %41, oturum bakımından %22 olarak gerçekleşmiştir. Tablet olarak adlandırılan çoğu zaman mobil bağlantıya sahip olmayan ve kablosuz veya kablolu olarak internete erişimi olan cihazlardan bağlantı oranı ise kullanıcı bakımından %1.78, oturum bakımından %1.31 olarak gerçekleşmiştir.

Bilgisayar, telefon, tablet ve benzeri cihazların ekranlarında bir pikselin gösteriminde kaç bit kullanıldığının bilgisi ekran bit derinliği olarak adlandırılır. Bit derinliği arttıkça ekrandaki

| | İşletim Sistemi | Yeni Kullanıcılar | Oturum |
|---|---|---|---|
| | | 1.927.495 | 8.568.575 |
| 1. | Android | 555.218 (%28,81) | 3.975.375 (%46,39) |
| 2. | iOS | 540.380 (%28,04) | 2.520.178 (%29,41) |
| 3. | Windows | 774.278 (%40,17) | 1.919.408 (%22,40) |
| 4. | Windows Phone | 8.467 (%0,44) | 41.304 (%0,48) |
| 5. | Macintosh | 14.241 (%0,74) | 31.378 (%0,37) |
| 6. | (not set) | 5.611 (%0,29) | 30.887 (%0,36) |
| 7. | SymbianOS | 22.628 (%1,17) | 24.368 (%0,28) |
| 8. | Linux | 4.054 (%0,21) | 14.844 (%0,17) |
| 9. | BlackBerry | 809 (%0,04) | 6.880 (%0,08) |
| 10. | Samsung | 548 (%0,03) | 1.396 (%0,02) |

**Şekil 10:** İşletim sistemine göre ziyaretçi sayıları.

| | Cihaz Kategorisi | Yeni Kullanıcılar | Oturum |
|---|---|---|---|
| | | 1.927.495 | 8.568.575 |
| 1. | mobile | 1.100.675 (%57,10) | 6.491.509 (%75,76) |
| 2. | desktop | 792.513 (%41,12) | 1.964.597 (%22,93) |
| 3. | tablet | 34.307 (%1,78) | 112.469 (%1,31) |

**Şekil 11:** Siteye erişilen cihazın kategorisine göre ziyaretçi sayıları.

görüntünün renkleri daha detaylı görüntülenebilir (Sullivan et al., 2012; Ohm et al., 2012; Duenas et al., 2012). Şekil 12'de görülebileceği üzere toplam oturumların yaklaşık %68'i yeni oturumlarınsa yaklaşık %56'sının ekran renk derinliğinin 32-bit olduğu görülmektedir. Oturumların yaklaşık %30'u ve yeni kullanıcıların yaklaşık %41'inin bit derinliği 24-bit olarak kaydedilmiştir. Bu iki ekran bit derinliği toplam ziyaretçilerin yaklaşık %95 gibi büyük bir kısmı yüksek bit derinliğinde siteyi ziyaret etmişlerdir.

Şekil 13'te siteye erişen kullanıcıların ekran çözünürlüklerinin oranları ve sayıları verilmiştir. Bu tabloda verilen bilgiler yatay ve dikey olarak ayrı ayrı değerlendirildiğinde 300-450 piksel genişlikteki çözünürlükler ile 480-700 piksel boyundaki çözünürlüklerin en büyük orana sahip oldukları görütülenebilir.

Bu çalışmanın örneklem zaman aralığında Türkiye'de satılan cep telefonu markalarını incelendiğinde 2010 yılında Nokia'nın %67 pazar payı ile lider konumda olduğu görülmektedir. 2019 yılına gelindiğinde ise Nokia'nın Pazar payı yüzde 0.62'ye düşmüştür. 2013 yılında ise Nokia piyasadaki pazar payını Samsung® ve Apple® şirketlerine bırakmıştır. Samsung® 2012 yılında %28 pazar payına sahipken 2013 yılında büyük bir sıçrama yaparak %45 pazar payına ulaşmıştır. 2019 yılında ise Samsung® %53 pazar payına sahip hâle gelmiştir. Apple®'ın ülkemizdeki pazar payı %17 ile %19 arasında değişkenlik göstermiştir (Ayan, 2019).

Siteye erişim sağlanan mobil cihazlar bağlamında incelemelerin yer aldığı Şekil 14'den görülebileceği üzere mobil cihazlar arasında oturum bakımından %37, yeni kullanıcı bakımından %45 oranla Apple Iphone birinci sırada yer almaktadır. Listenin devamında ise Samsung® marka cihazların olduğu görülebilmektedir. Burada Apple Iphone ®tüm modellerinin tek bir başlık altında listelenmesi ve diğer cihazların model bazlı dağılmış olması oranlar arasında farklılıkları daha net ortaya koymuştur.

Şekil 15'te verilen bilgilere göre servis sağlayıcılar bakımından Avea İletişim Hizmetlerini A.Ş.'nin toplam oturum sayısı bakımından yaklaşık %15, yeni kullanıcı bakımından %10 gibi bir orana sahip olduğu görülmektedir. Mobil servis sağlayıcılardan Vodafone toplam oturumda yaklaşık %10, Turkcell ise yaklaşık %9 ile listede yer almıştır. Listenin alt sıralarında ise sabit internet hizmetlerine ait TTnet ADSL hizmeti yer almaktadır.

Kullanıcılar web sitesine çeşitli kanallar vasıtası ile erişmektedir. Bu kanalları Şekil 16'daki gibi grupladığımızda toplam oturum bakımından yaklaşık %74 gibi bir oranda Organik Arama yani arama motorları üzerinden çeşitli anahtar kelimeler

| Ekran Renkleri | Yeni Kullanıcılar | Oturum |
|---|---|---|
| | 1.927.495 | 8.568.575 |
| 1. 32-bit | 1.089.126 (%56,50) | 5.844.431 (%68,21) |
| 2. 24-bit | 795.158 (%41,25) | 2.620.784 (%30,59) |
| 3. 16-bit | 30.914 (%1,60) | 52.534 (%0,61) |
| 4. 4-bit | 7.530 (%0,39) | 45.857 (%0,54) |
| 5. (not set) | 4.303 (%0,22) | 4.313 (%0,05) |
| 6. 0-bit | 253 (%0,01) | 370 (%0,00) |
| 7. 16777216-bit | 186 (%0,01) | 260 (%0,00) |
| 8. 8-bit | 14 (%0,00) | 15 (%0,00) |
| 9. 15-bit | 10 (%0,00) | 10 (%0,00) |
| 10. 24 bit | 1 (%0,00) | 1 (%0,00) |

**Şekil 12:** Cihazın ekran renklerine göre ziyaretçi sayıları.

| Ekran Çözünürlüğü | Yeni Kullanıcılar | Oturum |
|---|---|---|
| | 1.927.495 | 8.568.575 |
| 1. 360x640 | 385.640 (%20,01) | 2.656.095 (%31,00) |
| 2. 375x667 | 228.090 (%11,83) | 1.132.721 (%13,22) |
| 3. 1366x768 | 410.257 (%21,28) | 1.124.754 (%13,13) |
| 4. 320x568 | 172.927 (%8,97) | 812.016 (%9,48) |
| 5. 414x736 | 63.936 (%3,32) | 279.826 (%3,27) |
| 6. 320x480 | 55.381 (%2,87) | 242.710 (%2,83) |
| 7. 720x1280 | 25.364 (%1,32) | 202.215 (%2,36) |
| 8. 320x534 | 30.804 (%1,60) | 194.670 (%2,27) |
| 9. 412x732 | 17.123 (%0,89) | 155.107 (%1,81) |
| 10. 1920x1080 | 65.990 (%3,42) | 154.045 (%1,80) |

**Şekil 13:** Cihazın ekran çözünürlüğüne göre ziyaretçi sayıları.

| Mobil Cihaz Bilgileri | Yeni Kullanıcılar | Oturum |
|---|---|---|
| | 1.134.982 | 6.603.978 |
| | Toplam Yüzdesi: %58,98 (1.924.207) | Toplam Yüzdesi: %77,07 (8.568.575) |
| 1. Apple iPhone | 516.482 (%45,51) | 2.449.762 (%37,10) |
| 2. (not set) | 45.784 (%4,03) | 239.898 (%3,63) |
| 3. Samsung GT-I9500 Galaxy S IV | 27.369 (%2,41) | 205.172 (%3,11) |
| 4. Samsung GT-I8190 Galaxy S III Mini | 15.451 (%1,36) | 111.182 (%1,68) |
| 5. LG D855 G3 | 16.008 (%1,41) | 102.895 (%1,56) |
| 6. DOOGEE DG500 Discovery | 13.111 (%1,16) | 99.743 (%1,51) |
| 7. Samsung GT-N7100 Galaxy Note II | 14.257 (%1,26) | 99.697 (%1,51) |
| 8. Samsung GT-I9190 Galaxy S4 Mini | 14.407 (%1,27) | 99.248 (%1,50) |
| 9. Samsung SM-N9000Q Galaxy Note 3 | 10.993 (%0,97) | 95.940 (%1,45) |
| 10. Samsung GT-I9300 Galaxy S III | 10.618 (%0,94) | 84.280 (%1,28) |

**Şekil 14:** Erişimin sağlandığı mobil cihaza göre sayıları.

| Servis Sağlayıcı | Yeni Kullanıcılar | Oturum |
|---|---|---|
| | 1.927.495 | 8.568.575 |
| | Toplam Yüzdesi: %100,17 (1.924.207) | Toplam Yüzdesi: %100,00 (8.568.575) |
| 1. avea iletisim hizmetleri a.s. | 209.612 (%10,87) | 1.300.614 (%15,18) |
| 2. vodafone turkey 3g ip pool | 147.751 (%7,67) | 904.623 (%10,56) |
| 3. turkcell internet | 143.884 (%7,46) | 783.635 (%9,15) |
| 4. tt adsl-ttnet _dynamic_aci | 98.781 (%5,12) | 437.533 (%5,11) |
| 5. avea iletisim hizmetleri a.s | 65.302 (%3,39) | 437.263 (%5,10) |
| 6. tt adsl-tt net_aci | 82.213 (%4,27) | 385.843 (%4,50) |
| 7. tt adsl-tt net_dynamic_ulus | 41.950 (%2,18) | 216.052 (%2,52) |
| 8. tt adsl-ttnet_dynamic_aci | 52.324 (%2,71) | 201.518 (%2,35) |
| 9. superonline iletisim hizmetleri a.s. | 40.049 (%2,08) | 200.529 (%2,34) |
| 10. turk telekomunikasyon anonim sirketi | 38.451 (%1,99) | 187.188 (%2,18) |

**Şekil 15:** Ziyaret için kullanılan servis sağlayıcıya göre sayıları.

vasıtası ile arama yaparak çıkan sonuçlar üzerinden ulaşıldığı gözlemlenmiştir. Direkt olarak web sitesi adresini tarayıcının adres satırına yazarak giriş yapanların oranı ise toplam oturum bakımından %23, yeni kullanıcılar bakımından yaklaşık %33 olarak gerçekleşmiştir. Herhangi bir site üzerinden referans ile tıklama yaparak siteye ulaşanların oranı ise sırası ile %0.31 ve %0.56 olarak gerçekleşmiştir. Öte yandan sosyal medya siteleri üzerindeki paylaşımlarda bulunan bağlantıları tıklayarak siteye ulaşanların oranı ise %2.36 ve %1.73 olarak gerçekleşmiştir.

Ziyaretçilerin sitede hangi sayfaları ziyaret ettiği önemli bir bilgidir. Bu şekilde birçok faydalı sonuca ulaşmak mümkündür. Şekil 17'de verilen bilgilerde görüldüğü üzere toplam sayfa görüntülemelerinin yaklaşık %16'sında web sitesinin ana sayfası görüntülenmiştir. Ara sınav, yarıyıl sonu sınavı ve bütünleme sınav programları ile ders programlarının yayınlandığı sayfa ile öğrenci devam kontrol sistemi sayfalarının yoğun şekilde ziyaret edildiği görülmektedir. Sınav programları ise sırasıyla ara sınav programı %5.73 (%4.24 + %1.49), final sınav programı %5.28 (%3.92 + %1.36) ve bütünleme sınav programı %1.56 oranında ziyaret edilmiştir. Ders programı ise %9.49 (%7.22 + %2.27) oranında ziyaret edilmiştir.

Kullanıcı, internet tarayıcısının adres satırına ilgili web sitesinin adresini yazıp yükle dediğinde web sitesinin içerdiği metin ve görseller "Hiper Metin İşaret Dili" (Hyper Text Markup Language veya kısaca html ) formatında ağ üzerinden gönderilir ve kullanıcının tarayıcısı bu dili algılayarak görsel hâle getirerek kullanıcıya gösterir. Bu işlem çok kısa süre zarfında gerçekleşmektedir. Bu işlemin gerçekleşme süresini etkileyen bir dizi etmen bulunmaktadır. Bunların başında kullanıcının ve web sitesinin saklandığı internet sunucusunun internet hızları, sunucunun donanım özellikleri, ilgili web sayfasının taşıdığı görsel ve metin çeşitliği boyutu vb. gelmektedir. Bu ve benzeri etkenler sayfanın yüklenmesi için gerekli süreyi belirlemektedir. Sayfanın uzun sürede açılması kullanıcıda siteye erişimden vazgeçmeye kadar giden bir davranışa yol açabilmektedir. Günümüz dünyasında teknolojik gelişmeler ile kablolu ve kablosuz bağlantı hızlarının tatmin edici düzeylere ulaştığı gözlemlendiğinde sayfaların açılma hızlarının çok ciddi oranlarda düşüşe uğrayacağı öngörülebilmektedir.

Şekil 18'de ilgili tarihler arasında yükseköğretim kurumu web sitesinin ortalama süreleri ile ilgili bilgiler yer almaktadır. Ortalama sayfa yükleme süresi 2.62 saniye, ortalama yönlendirme süresi 0.1 saniye, ortalama alan arama süresi 0.03 saniye, ortalama sunucu bağlantı süresi 0.11 saniye, ortalama sunucu yanıt süresi 0.48 saniye ve ortalama sayfa indirme süresi 0.18 saniye olarak gerçekleşmiştir.

Şekil 19'da ise ilgili tarihler arasından yükseköğretim kurumu web sitesinin tarayıcı bazlı ortalama sayfa yüklenme süreleri ile ilgili bilgiler yer almaktadır. İnternet tarayıcısı ortalama sayfa yükleme sürelerini incelediğimizde 0.41 saniye ile Microsoft Edge, 0.72 sn ile Mozilla Firefox ve 0.78 saniye ile YaBrowser en hızlı sayfa yükleme ortalamasına sahip tarayıcılardır. Diğer tarayıcıların ortalama süreleri Şekil 19'da detaylı olarak görülebilir.

Şekil 20'de ülke bazlı ortalama sayfa yükleme süreleri görülebilmektedir. Buna göre 9.13 saniye ortalama ile en uzun sayfa yükleme ortalamasına sahip Amerika Birleşik Devletleri

| Default Channel Grouping | Yeni Kullanıcılar ⓘ | Oturum ⓘ |
|---|---|---|
| | **1.927.495** Toplam Yüzdesi: %100,17 (1.924.207) | **8.568.575** Toplam Yüzdesi: %100,00 (8.568.575) |
| 1. (Diğer) | 1 (%0,00) | 1 (%0,00) |
| 2. Direk | 644.387 (%33,43) | 1.981.536 (%23,13) |
| 3. Organik Arama | 1.238.998 (%64,28) | 6.358.020 (%74,20) |
| 4. Referans | 10.727 (%0,56) | 26.889 (%0,31) |
| 5. Sosyal Medya | 33.382 (%1,73) | 202.129 (%2,36) |

**Şekil 16:** Site erişim kanalları.

| Sayfa ⓘ | Sayfa Görüntüleme Sayısı ⓘ ↓ | Farklı Sayfa Görüntüleme Sayısı ⓘ | Sayfada Geçirilen Ortalama Süre ⓘ | Giriş Sayısı ⓘ | Hemen Çıkma Oranı ⓘ | Çıkış Yüzdesi ⓘ |
|---|---|---|---|---|---|---|
| | **28.220.548** | **22.002.698** | **00:01:02** | **8.568.447** | **%35,94** | **%30,36** |
| 1. /tr/ | 4.414.842 (%15,64) | 3.463.081 (%15,74) | 00:00:52 | 3.162.069 (%36,90) | %26,50 | %26,86 |
| 2. /wp-courselist.php | 2.326.753 (%8,24) | 1.587.774 (%7,22) | 00:00:27 | 755.926 (%8,82) | %9,18 | %11,17 |
| 3. /wp-exam-schedule-ara.php | 1.294.965 (%4,59) | 933.077 (%4,24) | 00:00:24 | 382.788 (%4,47) | %7,14 | %9,54 |
| 4. /wp-exam-schedule-fin.php | 1.170.213 (%4,15) | 863.082 (%3,92) | 00:00:22 | 369.712 (%4,31) | %6,85 | %10,29 |
| 5. /bolumlere-ve-derslere-gore-ders-programi/224/ | 576.115 (%2,04) | 499.476 (%2,27) | 00:00:05 | 169.864 (%1,98) | %6,99 | %4,63 |
| 6. /wp-exam-schedule-but.php | 490.350 (%1,74) | 344.277 (%1,56) | 00:00:23 | 130.240 (%1,52) | %7,05 | %10,74 |
| 7. /wp-att/index.php | 466.873 (%1,65) | 194.068 (%0,88) | 00:01:06 | 31.379 (%0,37) | %50,04 | %32,78 |
| 8. /wp-course-catalog.php | 407.557 (%1,44) | 140.181 (%0,64) | 00:01:29 | 36.159 (%0,42) | %32,44 | %25,22 |
| 9. /ara-sinav-programi/2956/ | 365.611 (%1,30) | 327.929 (%1,49) | 00:00:05 | 78.874 (%0,92) | %8,11 | %4,04 |
| 10. / | 361.075 (%1,28) | 313.797 (%1,43) | 00:00:57 | 259.317 (%3,03) | %15,90 | %17,27 |
| 11. /final-sinavi-programi/2984/ | 331.687 (%1,18) | 299.162 (%1,36) | 00:00:04 | 68.533 (%0,80) | %8,50 | %3,96 |
| 12. /studentdash.php | 324.570 (%1,15) | 262.215 (%1,19) | 00:00:15 | 898 (%0,01) | %9,69 | %9,84 |
| 13. /akademik-takvim/211/ | 298.567 (%1,06) | 245.193 (%1,11) | 00:02:08 | 121.064 (%1,41) | %69,63 | %57,60 |

**Şekil 17:** Sayfa bazlı ziyaret ve istatistik bilgisi.

olmuştur. Bu ülkeyi sırası ile 8.03 sn ile Bulgaristan, 4.66 ile Bosna Hersek ve 4.21 ile Hollanda izlemektedir. Türkiye 1.57 sn, Almanya 1.53 sn ve Kıbrıs adasından 1.43 sn. ortalama sayfa yükleme süreleri ile sitenin genel yükleme süresi ortalamasının altında yükleme süresi ortalamasına sahiptirler.

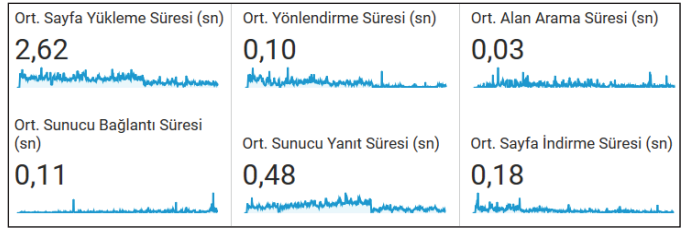

**Şekil 18:** Ortalama süreler.

| | Tarayıcı | Ort. Sayfa Yükleme Süresi (sn) |
|---|---|---|
| 1. | Edge | 0.41 |
| 2. | Firefox | 0.72 |
| 3. | YaBrowser | 0.78 |
| 4. | Safari (in-app) | 1.20 |
| 5. | Samsung Internet | 1.33 |
| 6. | Android Webview | 1.47 |
| 7. | Safari | 1.57 |
| 8. | Chrome | 1.61 |
| 9. | Opera | 2.01 |
| 10. | Internet Explorer | 2.58 |

**Şekil 19:** İnternet tarayıcısı bazlı ortalama sayfa yükleme süreleri.

## DEĞERLENDİRMELER

Çalışmada 2013 ile 2019 yılları arasındaki ziyaretçi verileri kullanılmıştır. İlgili dönemde web sitesi istatistiklerine göre 2 milyona yakın tekil kullanıcı ziyareti ve 8.5 milyonun üstünde oturum gerçekleşmiştir. Her bir kullanıcı yaklaşık 4.5 oturum ile siteye erişmiştir. Buradaki oturum bir kullanıcının farklı zamanlarda siteyi ziyaret etmesi anlamına gelmektedir. Bir kullanıcının sitede ortalama olarak 2 dakika 23 saniye kaldığı hesaplanmıştır. Siteye giren 100 kullanıcında yaklaşık 36 adetinin ise hemen çıktığı görülebilmektedir. Sitenin ziyaretçilerinin yaklaşık %70'i yeni kullanıcılar, geriye kalan %29'u ise daha önce siteyi ziyaret edip tekrar siteye giren kullanıcılardan oluşmaktadır. Tekil kullanıcı sayısı temel alındığında iki milyon farklı kullanıcının siteye erişmesi ciddi bir sayıdır. Belli bir zümreyi ilgilendiren yükseköğretim kurumu için iki milyon gibi bir ülke nüfusuna eşit olabilecek boyutta tekil kullanıcı çekmesi içerik ve ulaşılabilirlik konusunda sitenin iyi olduğu anlamına gelmektedir. Ayrıca sitede geçirilen ortalama süre ve her bir tekil kullanıcının oturum sayıları bakımından da sitenin kullanıcıları içeriği ve tasarımı ile sıkmadığı ve site içinde tutabildiği sonuçları çıkarılabilir. Siteyi ziyaret eden kullanıcıların yaklaşık %70'inin yeni kullanıcı olması ziyaretçilerin kalıcı olmadığı anlamını taşıyor gibi görünse de %30 ziyaretçi sayıları göz önünde bulundurulduğunda yaklaşık 600 bin tekil kullanıcının siteyi tekrar ziyaret ettiği ve dönüş sağladığı anlamını taşımaktadır ki bu ciddi bir ziyaretçi kapasitesidir. İlgili ziyaretçi sayıları ve ziyaretçilerin davranışları neticesinde sitenin yazılım ve sunucu kaynak ihtiyaçları planlanmalı, kullanıcı ihtiyaçlarına göre içerik düzenlenmelidir. Sunucu cevap sürelerini azaltacak önlemler alınmalıdır.

İlgili tarih aralığında günlük olarak ziyaretçi trafiği izlenmiştir. 6 yıllık süre zarfında 5000 kullanıcı ve üzerine çıkan gün sayısının aylara göre dağılımları sırasıyla Haziran, Ocak ve Ekim ayları ön plana çıkmaktadır. Bu üç ayın yoğun olarak ziyaretçi çekmesinin nedeni Akademik takvim göz önünde bulundurularak irdelendiğinde ilgili tarihlerin dönem başı ve sonuna denk geldiği

| | Ülke | Ort. Sayfa Yükleme Süresi (sn) | Ort. Sayfa Yükleme Süresi (sn) (site ortalamasıyla karşılaştırıldığında) |
|---|---|---|---|
| | | 1,58 | 1,58 |
| 1. | United States | 9,13 | %476,02 |
| 2. | Bulgaria | 8,03 | %407,08 |
| 3. | Bosnia & Herzegovina | 4,66 | %194,24 |
| 4. | Netherlands | 4,21 | %165,89 |
| 5. | Czechia | 3,59 | %126,74 |
| 6. | (not set) | 2,34 | %47,46 |
| 7. | Serbia | 2,20 | %38,85 |
| 8. | Turkey | 1,57 | -%0,82 |
| 9. | Germany | 1,53 | -%3,41 |
| 10. | Cyprus | 1,43 | -%10,00 |

**Şekil 20:** Ülke bazlı sayfa ortalama sayfa yüklenme süreleri.

izlenebilir. Ekim ayı; yaz okulu bitiş ve güz dönemi başlangıcı, Ocak ayı; güz dönemi bitiş bahar yarıyılı başlangıcı ve Haziran ayı; bahar yarıyılı bitiş ve yaz okulu başlangıç dönemlerine denk gelmektedir. Bu süreler zarfında sitenin yoğun trafik aldığı düşünüldüğünden gerekli teknik önlemlerin alınması ve sitenin hizmet vermesinin aksamaması için ilgili çalışmaların yapılması önem taşımaktadır.

İstatistikler incelendiğinde toplam ziyaretçilerin %55'inin erkek, %45'inin kadın olduğu görülmektedir. Aynı tarih aralığındaki toplam ziyaretçilerin %27.5'i 18-24 yaş, %33.5'i 25-34 yaş, %15.5'i 35-44 yaş ve geri kalan %23.5'lik kısmı ise 45 ve üstü kullanıcılardan oluşmaktadır. Kısacası kullanıcıların %61'i 35 yaş altındadır. Buradan da anlaşılacağı gibi yükseköğretim kurumunda 18-24 yaş grubundaki öğrencilerin eğitim öğretim aldıkları ve mezun olanların erişim durumları da dikkate alındığında kullanıcıların genç bireylerden olduğu anlaşılmaktadır. Bu bireylerin yüksek lisans veya doktora yapma ihtimalleri, yeni mezun olmaları nedeni ile aidiyet duygusunun güçlü olması, teknoloji kullanımının genç bireylerde daha yüksek olması ve aktif öğrencilerin ziyaretinin çok olması gibi nedenler bu oranların oluşumuna neden olmuş olabilir. Ayrıca genç yaş grubundan ziyaretçilerin sayısının yüksek olması ilgili yükseköğretim kurumu ile doğrudan ilişkili olmayan diğer yükseköğretim kurumlarından veya ilgi duyan diğer gençlerin de ziyaret ettiği anlamı çıkarılabilir. Bu durumlarda göz önünde bulundurularak öncelikle genç ziyaretçilerin talebini canlı tutmak ve diğer yaş gruplarından ziyaretçilere hitap edecek çeşitli etkinlik ve ilgilerini çekebilecek içerikler eklenmesi uygun olacaktır. Kadın bireylerin ziyareti yükseköğretim kurumunun mezun ve okuyan öğrenci profilindeki cinsiyet oranı ile ilgili olabileceği gibi kadınların ilgisinin azaldığı anlamı da çıkarılabilir.

Siteye erişimler oturumlar bakımından değerlendirildiğinde yaklaşık %95 'inin Türkçe dili ile yaklaşık %4.5 İngilizce dili ve geri kalan kısmını Rusça, Almanca gibi diller oluşturmaktadır.

Türkiye %98 ile birinci sırada yer alırken onu Amerika Birleşik Devletleri, Rusya, Almanya ve Birleşik Krallık izlemektedir. Toplam oturumun yaklaşık %98'i Asya, %64'ü Avrupa ve diğer kısımları Amerika, Afrika ve Avustralya kıtası olmak üzere sıralanmıştır. Ziyaretçiler şehirler bakımından incelendiğinde ise en çok ziyaretçi %58 tekil ziyaretçi ve %74 oturum sayısı ile İzmir'den gerçekleşmiştir. Oturum sayısı bakımından bu şehri %9.36 ile İstanbul, %4.51 ile Ankara ve %1.08 ile Antalya izlemektedir. Oturum sayısı bakımından genel sıralamada 26. sırada yer alan New Delhi %0.12'lik oturum oranı ile yurtdışı şehirleri arasında birinci sırada yer almaktadır. Onu %0.08 ile Moskova ve %0.05 ile Londra izlemektedir. Bu istatistiklerden anlaşılacağı üzere site ağırlıklı olarak ülkemizden ziyaretçi çekmektedir. Sitenin tüm kaynağının İngilizcesinin de olduğu göz önünde bulundurulduğunda ilgili içeriğin yurtdışından ziyaretçi çekmesi için yeterli ve ilgi çekici olmadığı, bunun bilinirliğin düşük olduğu ve uluslararası öğrenci veya paydaşların düşük olduğu sonucu çıkarılabilir. Bu alanda uluslararası tanıtım, içerik üretme ve mevcut içeriklerinin siteye eklenerek uluslararası ziyaretçilerin siteye daha büyük oranda erişmesi sağlanabilir.

Sayfa erişimleri internet tarayıcısı bakımından değerlendirildiğinde ise kullanıcıların yaklaşık %58'i ve oturumların yaklaşık %59'unun Chrome tarayıcısını kullandığı görülmektedir. Bu tarayıcıyı yaklaşık %25 ile Safari izlemektedir. Diğer tarayıcılar ise Andorid browser, Internet Explorer ve Firefox vb. şeklindedir. İşletim sistemi bakımından ise oturumların %46, kullanıcıların ise %28'inin Android, %28 kullanıcı ve %29 oturum oranı ile IOS işletim sisteminden gerçekleştiği görülebilmektedir. İşletim sistemi bakımından ise üçüncü sırada ise %40 kullanıcı, %22 oturum oranı ile Windows işletim sistemi yer almaktadır. Bu bilgiler bize kullanıcıların ağırlıklı olarak Android ve IOS işletim sistemini kullanan cihazlardan siteye eriştiğini ve tüm işletim sistemleri üzerinde sağlıklı çalışan Chrome tarayıcısının en çok kullanılan tarayıcı olduğunu göstermektedir. Site üzerinde yapılacak değişikliklerin, mobil uygulamaların bu bilgiler değerlendirilerek yapılması önem kazanmaktadır.

Erişim sağlanan cihazların türüne göre gruplama yapıldığında ise cep telefonu vb. (mobil cihazlar) üzerinden erişim oranı kullanıcı bakımından %57, oturum bakımından %76 olarak gerçekleşmiştir. Masaüstü olarak tabir edilen bilgisayarlardan erişim oranı kullanıcı bakımından %41, oturum bakımından %22 olarak gerçekleşmiştir. Ziyaretçilerin kullandığı cihazların renk derinliği dikkate alındığında oturumların yaklaşık %68'i yeni oturumların ise yaklaşık %56'sının ekran renk derinliğinin 32-bit olduğu görülmektedir. Oturumların yaklaşık %30'u ve yeni oturumların yaklaşık %41'inin bit renk derinliği ile 24-bit olarak kaydedilmiştir. Erişen cihazların ekran çözünürlüğü bakımından 300-450 piksel genişlikteki çözünürlükler ile 480-700 piksel boyundaki çözünürlüklerin tüm ziyaretçilerin büyük bir bölümünü oluşturduğu tespit edilmiştir. Bu bilgilerden anlaşılacağı üzere masaüstü bilgisayarlardan ve işletim sistemlerinden erişimler günden güne azalmakta ve mobil cihazların kullanımının yaygınlaşması ile erişim oranları o yönde artış göstermektedir. Tasarımların mobil uyumlu hâle getirilmesi, farklı çözünürlüklerde sorunsuz çalışan tepkisel arayüzler geliştirilmesi önem kazanmıştır. Renk derinlikleri bakımından 24 ve 32 bit derinlikler tatmin edici sonuçlar verdiğinden sitedeki görseller ve içerikler cihazlarda okunabilir ve tatmin edici görünümü sağlayabilir durumdadır denilebilir. Ayrıca tüm bunlara ek olarak mobil cihazların desteklediği ve desteklemediği teknolojiler site geliştirme aşamalarında dikkate alınmalıdır.

Sayfa erişim istatistikleri erişim sağlanan mobil cihazlar bakımından incelendiğinde Apple Iphone® marka cihazların birinci sırayı aldığı görülmektedir. Listenin devamında ise Samsung® marka birçok modelde cihazın yer aldığı tespit edilmiştir. Marka bakımından liste üzerinden okuma yapıldığında Iphone %38.16 ile birinci sırada, Samsung® %30.43 ile ikinci sırada ve LG %6.16 oranla üçüncü sırada yer almaktadır. Buradan anlaşılacağı üzere genç bir kitlenin ziyaret ettiği bu sitenin kullanıcılarının mobil cihaz kullanım oranları, cihazların Türkiye'deki satılma oranları ile paralellik göstermemektedir. Satılma oranlarında ilk sırada Samsung® yer almasına rağmen erişim istatistiklerinde ilk sırada Iphone yer almaktadır. Iphone® kullanımının genç bireylerde daha çok olduğu buradan çıkarılabilir.

Erişim sağlayıcısı operatörler bakımından incelendiğinde toplam oturumların yaklaşık %15'i, tekil kullanıcıların %10'u Avea operatörü ile siteye erişmiştir. Mobil servis sağlayıcılarından Vodafone ise ikinci sırada yer almaktadır. Turkcell ise listede üçüncü sırada yer almıştır. Buradan anlaşılacağı üzere Avea operatörü siteye erişen kullanıcılar arasında daha büyük bir kullanım oranına sahiptir. Avea'nın uyguladığı fiyat politikası, hizmet kalitesi ve kampanyaları erişim sağlayanların genç bir kitle olduğu göz önünde bulundurulduğunda daha cazip geldiği söylenebilir.

Siteye erişim kanalları bakımından bulgulara bakıldığından arama motorları vb. aracılığı ile siteye ulaşanların oranı %74 gibi çok büyük bir orana sahiptir. Tarayıcı adres satırına ilgili web site adresini yazarak direkt erişen kullanıcı oranı ise %24 seviyesinde gerçekleşmiştir. Sosyal medya siteleri üzerinde yer alan paylaşımlardan yapılan tıklamalar ile ulaşım oranı %2.3 olarak gerçekleşmişken, sosyal medya dışındaki web siteleri üzerinden referansla ulaşanların oranı ise %0.4 olarak gerçekleşmiştir. Bu istatistiklerden anlaşılacağı üzere site arama motorlarında üst sıralarda çıkmaktadır. Ziyaretçiler çoğunlukla adresi akılda tutarak giriş yapmak yerine arama motorlarından aramayı daha kolay ve erişilebilir bulmaktadır. Bunun bir nedeni de sitenin ana sayfasına değil de devam kontrol sistemi, ders programı, sınav programı gibi alanlara erişimde direkt bağlantıya arama motorları vasıtası ile ulaşıldığından kaynaklanmaktadır. Sayfa bazlı trafik bilgisi analizi bu sonuçları bize göstermektedir.

İnternet tarayıcısı ortalama sayfa yükleme sürelerini incelediğimizde 0.41 saniye ile Microsoft Edge, 0.72 sn ile Mozilla Firefox ve 0.78 saniye ile 0.78 saniye ile YaBrowser en hızlı sayfa yükleme ortalamasına sahip tarayıcılardır. Öte yandan ülkeler bakımından değerlendirildiğinde 9.13 saniye ortalama ile en uzun sayfa yükleme ortalamasına sahip Amerika Birleşik Devletleri olmuştur. Bu ülkeyi sırası ile 8.03 sn ile Bulgaristan, 4.66 ile Bosna Hersek ve 4.21 ile Hollanda izlemektedir. Türkiye 1.57 sn ortalama ile sitenin genel yükleme süresi ortalamasının altında bir sürede yükleme süresi ortalamasına sahiptir. Özellikle Balkan ülkelerindeki erişim yavaşlığının ilgili ülke internet altyapısından kaynaklanabileceği öngörülmektedir. Tüm bunlara ek olarak sayfa açılma sürelerini etkileyen resim boyutları, javascript vb. hatalı kodlar ve html kod boyutu optimizasyonu gibi çözümlere başvurulması önerilmektedir.

### SONUÇ

Çalışma ile ülkemizde alanında önde gelen bir yükseköğretim kurumunun web sitesi ziyaretçi verisi incelenmiştir. 2013 ve 2019 gibi 6 yıllık uzun bir süre için ziyaretçi trafiğinin izlenmesi ile elde edilen veriler çeşitli istatistiksel gruplamalar ile anlamlandırılmıştır. Web sitelerinin gelişmesi, tekeplere ve gereksinimlere göre şekillenmesi, günün ihtiyaçlarına cevap verilmesi açısından bu analizler büyük katkı sunmaktadır. Bu çalışma ile elde edilen veriler ve yapılan analizler neticesinde arama motorlarında ön sıralarında çıkma, çok çeşitli cihazlar üzerinde uyumlu bir şekilde çalışma, sayfa açılış sürelerinin optimize edilmesi, dönemsel kaynak ihtiyaçlarının belirlenmesi, kullanıcı tercihlerinin belirlenerek web sitesinin güncellenmesi, sorunların tespit edilip giderilmesi gibi birçok işlemi yerine getirme imkânı doğmuştur. Bu anlamda ilgili işlemlerin ve yapılan analizlerin toplum ile paylaşılması bu konuda ihtiyaç sahibi olan web sitelerine yol gösterici olacaktır. Bu verilerin paylaşılıyor olması ise kurumsal birçok web sitesinin kendi çalışmalarını yapma ve buradaki veriler ile karşılaştırma imkânını doğuracaktır.

### KAYNAKLAR

Arma Digital (2019). Web sitesi analizi. Retrieved from https://www.armadigital.net/web-tasarim-blog/web-sitesi-analizi

Ayan, S. (2019). İşte son on yılda cep telefonu satışlarının değişimi. Retrieved from https://www.tknlj.com/iste-son-on-yilda-cep-telefonusatislarinin-degisimi/

Businessht Bloomberght. (2015). Türkiye'de internetin hızlı tarihi. Retrieved from https://businessht.bloomberght.com/teknoloji/haber/1174426-turkiyede-internetin-hizli-tarihi

Dueñas, A., Malamy, A., Olofsson, B., Ichigaya, A., Sakaida, S., Pejhan, S., ... & Jones, T. (2012). On a 10-bit consumer-oriented profile in High Efficiency Video Coding (HEVC). ITU-T/ISO/IEC Joint Collaborative Team on Video Coding (JCT-VC) Document JCTVC-K0109.

Kavanagh, S. (2019). How fast is 5G? Retrieved from https://5g.co.uk/guides/how-fast-is-5g/

Küçük, Z. (2017). İnternetin gelişimi ve günümüzdeki durumu. Retrieved from https://womaneng.com/internetin-gelisimi-ve-gunumuzdekidurumu/

Ohm, J. R., Sullivan, G. J., Schwarz, H., Tan, T. K., & Wiegand, T. (2012). Comparison of the coding efficiency of video coding standards—including high efficiency video coding (HEVC). IEEE Transactions on circuits and systems for video technology, 22(12), 1669-1684.

Shiftdelete.net (2011). Fiber optik İle 100 tbps bağlantı. Retrieved from https://shiftdelete.net/fiber-optik-ile-100-tbps-baglanti-29021

Sohn, T., Li, K. A., Griswold, W. G., & Hollan, J. D. (2008). A diary study of mobile information needs. In Proceedings of the SIGCHI Conference on Human Factors in Computing Systems (pp. 433-442). ACM.

Sullivan, G. J., Ohm, J. R., Han, W. J., & Wiegand, T. (2012). Overview of the high efficiency video coding (HEVC) standard. IEEE Transactions on circuits and systems for video technology, 22(12), 1649-1668.

Wikipedia. (2018). Satellite internet access. Retrieved from https://en.wikipedia.org/wiki/Satellite_Internet_access